\documentclass[11pt,twoside]{article}

\usepackage{asp2006}
\usepackage{epsf}
\usepackage{psfig}
\usepackage{lscape}

\begin{document}
\title{Wide profile drifting pulsars : an elegant way to probe pulsar magnetospheres}   %%% Fill in title
\author{Bhaswati Bhattacharyya $^1$, Yashwant Gupta$^1$ and Janusz Gil$^2$} %%% Fill in author names
\affil{$^1$NCRA $-$ TIFR Pune, India  \\  %%% Fill in author affiliations
$^2$University of Zielona Gora, Poland}

\begin{abstract} %%% Abstract to run on from here.
We present the results from a study of wide profile pulsars using high
sensitivity multifrequency observations with the GMRT. Since the line of sight
samples a large region of the polar cap in case of the wide profile pulsars, presence
of simultaneous multiple drift regions is quite probable (as seen in PSR B0826$-$34 and
PSR B0818$-$41). We solve the aliasing problem of PSR B0818$-$41 using the observed phase
relationship of the drift regions, and determine its pattern rotation period $P_4$ to be 
$\sim$ 10s, which makes it the fastest known carousel. We find that, for all the pulsars 
showing drifting in multiple rings of emission, the drift pattern from the rings are 
phase locked.  This can constraint the theoretical models of pulsar emission 
as it favors a pan magnetospeheric radiation mechanism. 
\end{abstract}

%%% MAIN BODY OF TEXT GOES HERE. CONSULT "INSTRUCTIONS FOR AUTHORS USING
%%% LATEX2E MARKUP", SECTIONS 2.3-2.6 FOR HELP WITH EQUATIONS, FIGURES,
%%% AND TABLES.

\section{Introduction}   %%% Top level section head (remove "%" symbol)
Most pulsars have a narrow duty cycle of emission (5-10 \% of pulsar period). This is 
generally consistent with the expectations of the angular width of the polar cap, given 
typical viewing geometries.  However, there are a small but significant number of pulsars 
with unusually wide profiles, for which the emission is seen over a wide range of longitudes 
($\geq$ 90 degrees).  This is expected from pulsars which are highly aligned, i.e. the 
magnetic dipole axis is almost parallel to the spin axis, and emission is seen from 
one magnetic pole of the pulsar. In such a case, the line of sight (LOS) is very 
close to both the rotation and the magnetic axes, and consequently, we sample a large 
region of the polar cap. This has the exciting potential for a detailed study of the 
distribution and behavior of emission regions located in annular rings around the magnetic 
axis.  The study of pulsars showing systematic subpulse drift patterns provides important 
clues for understanding the unsolved problems of pulsar emission mechanism. Constraints 
from such observations can have far reaching implications for the theoretical models, as 
exemplified by some recent results (e.g. Deshpande \& Rankin (1999) and Gupta et al. (2004)).  
In this context, wide profile drifting pulsars can provide extra insights because of the 
presence of multiple drift bands.  In this work, we have mainly concentrated on studies 
of two such pulsars, PSR B0818$-$41 and PSR B0826$-$34.

B0818$-$41 is a relatively less studied wide profile pulsar with emission occurring for 
more than 180 deg of pulse longitude. It has a period of 0.545 $s$ and is relatively old, 
with a characteristic age of $4.57\times10^{8}$ years. The inferred dipolar magnetic field 
of this pulsar is $1.03\times10^{11}$ G, which is a typical value for slow pulsars.  From 
a study of its average polarization behaviour at 660 and 1440 MHz, Qiao et al. (1995) 
predict that the pulsar must have a small inclination angle between the magnetic and 
rotation axes.

PSR B0826$-$34 is a pulsar with one of the widest known profiles.  Earlier studies of 
this pulsar (Durdin et al. (1979), Biggs et al. (1985), Gupta et al. (2004)) have revealed 
some unique properties : strong evolution of the average profile with frequency, apparent 
nulling for 70\% of time, and a remarkable subpulse drift property $-$ multiple curved drift 
bands with frequent changes and sign reversals of drift rate.

\section{Observations and preliminary analysis}
The GMRT is a aperture synthesis telescope consisting of 30 antennas, each of 45 m diameter, 
spread over a region of 25 km diameter. The GMRT can also be used in the tied array mode by 
adding the signals from the individual dishes, either coherently or incoherently 
(Gupta et al 2000).  We performed polarimetric observations of PSR B0818$-$41 and 
PSR B0826$-$34 at multiple frequency bands (157, 244, 325, 610 and 1060 MHz), at different 
epochs, using the phased array mode of the GMRT.  Some of the observations were simultaneous 
at 303 and 610 MHz, and were carried out using the ``band masking'' technique, details of 
which are described in Bhattacharyya et al. (2008).  Data were recorded at a sampling rate 
of 0.512 ms.  During the off-line analysis the raw data were further integrated to achieve 
the final resolution of 2.048 ms.  

Average pulse profiles at different observing frequencies were obtained by dedispersion with 
a DM of 113.4 $pc/cm^3$ for PSR B0818$-$41, and 52.2 $pc/cm^3$ for PSR B0826$-$34.  The main 
source of corruption of the data was found to be power line interference (50 Hz and its 
harmonics).  The dedispersed data were put through a filtering routine which detected most 
(but probably not all) of the power line interferences and replaced them by appropriate 
random noise.  The dedispersed, interference free data stretch were then synchronously 
folded with the topocentric pulsar period.  Single pulse data streams were also generated 
from the dedispersed data, for both total intensity and full Stokes parameters.

\section{Results}
{\bf PSR B0818$-$41:} We confirm the remarkable subpulse drift pattern for this pulsar, first 
reported by us in Bhattacharyya et al. (2007). At 325 MHz, we can clearly see simultaneous 
occurrence of three drift regions with two different drift rates (Fig 1a): an inner region 
with steeper apparent drift rate, flanked on each side by a region of slower apparent drift 
rate.  Similar subpulse drifting is observed in both the inner and outer regions at 610 MHz, 
though the drift bands are weaker (Fig 1b).  The closely spaced drift bands always maintain 
a constant phase relationship: the subpulse emission from the inner drift region is in phase 
with that from the outer drift region on the right hand side, and at the same time the 
emission in the inner drift region is out of phase with the outer drift region situated on 
the left hand side.

A new technique is introduced by us for resolving aliasing, utilising the constant offset 
($\sim9P_1$) that is seen between the peak emission from the leading and trailing outer 
regions (Bhattacharyya et al. 2008b).  The basic idea here is that combination of the 
angular separation between the sparks in the outer ring, the time of traverse of the LOS
from the leading to the trailing outer component, and the drift rate of the sparks
in the outer ring should all be matched so as to produce that this offset.  From the results 
of our analysis, we find that the unaliased drift is too slow to allow this to happen.  We 
propose that the drift rate is most likely first order aliased, and the corresponding pattern 
rotation period, $P_4$, is 10 s.  This implies that PSR B0818$-$41 has the fastest known 
carousel.\\

{\bf PSR B0826$-$34:} At any given time, the simultaneous multiple subpulses present in the 
main pulse window follow the same drift rate and sign, at both the frequencies (Figs 2a \& 2b).  
The pulse regions showing different drift rates of opposite signs are always connected by 
a region showing smooth transition of drift rate. The gray scale plot of the single pulses 
from the higher sensitivity single frequency observations at 610 and 1060 MHz show coherent 
drifting in the main pulse (MP) and inter pulse (IP) regions, with approximately 6 drift 
bands in the MP and 4 drift bands in the IP, as seen in Figs 2a \& 2b 
(see also Bhattacharyya et al. 2008a).  The drifts in MP and IP regions are always locked 
in phase.

\begin{figure}
\plottwo{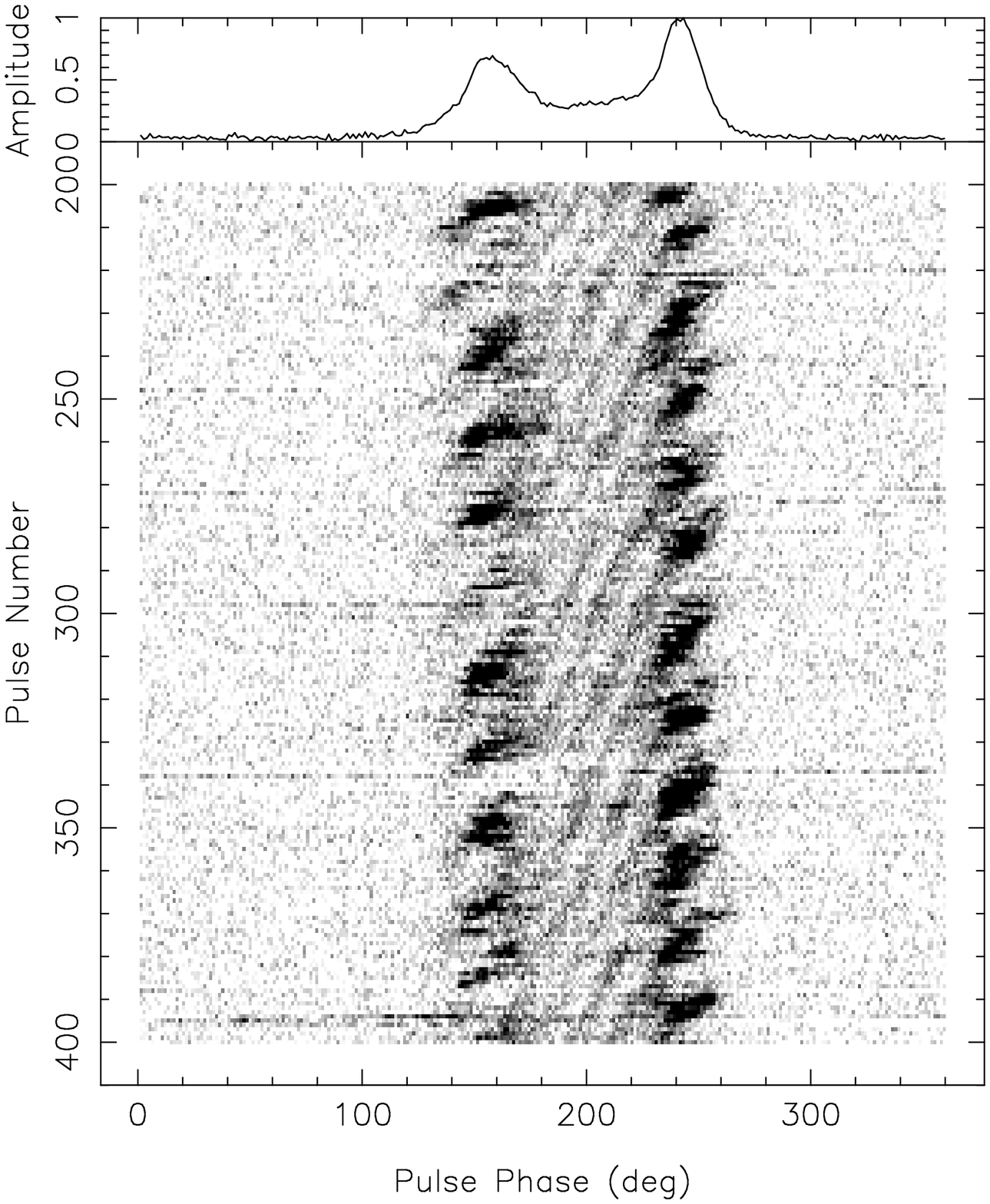}{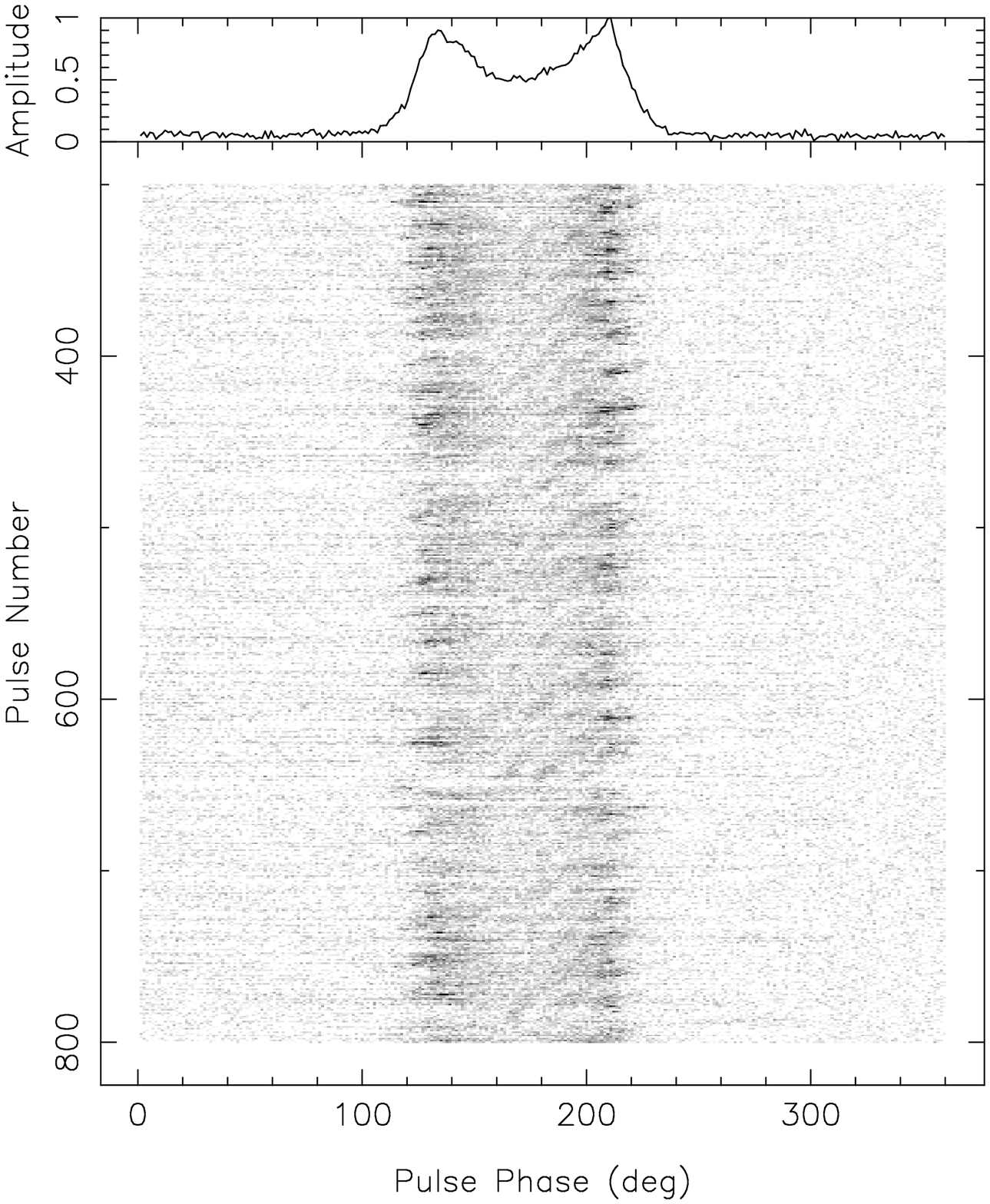}
\caption{{\itshape Left (Fig 1a) : Gray scale plot of the single pulse data from single 
frequency observations of PSR B0818$-$41 at 325 Hz.}
{\itshape Right (Fig 1b) : Same as Fig 1a, but at 610 Hz.}}
\end{figure}

\section{Interpretations}
{\bf PSR B0818$-$41:} The unique drift pattern of this pulsar can be naturally explained 
as being created by the intersection of our LOS with two conal rings on the polar cap of a 
fairly aligned rotator.  Based on the frequency evolution of the average profile, observed 
PA swing and results from subpulse drifting, we converged on two possible choices of 
emission geometry: {\bf G-1} ($\alpha=11$ deg and $\beta=-5.4$ deg) and {\bf G-2} 
($\alpha=175.4$ deg and $\beta=-6.9$ deg).  Whereas the features of the observed drift 
pattern are better reproduced by simulations of the radiation pattern using {\bf G-1}, 
the geometry {\bf G-2} appears to produce a better fit to the position angle swing of the
linear polarisation. 

Though the regular drift patterns observed in Fig. 1 are quite common for this pulsar, we 
sometimes observe changes of the drift rates, which are almost always associated with the 
occurrence of nulls.  However, the phase locked relation is maintained across the regions 
of irregular drifting or nulling.  This phase locked relation implies a common electrodynamic 
control between the rings.\\

{\bf PSR B0826$-$34:} The radiation pattern is interpreted to be created by the 
intersection of our LOS with two conal rings on the polar cap of a fairly aligned rotator 
(Gupta et al. (2004), Esamdin et al. (2005), Bhattacharyya et al. 2008a).  The observed wide 
pulse profile and its remarkable evolution with frequency can be explained by the intersection 
of the LOS with two concentric rings of emission.  At lower frequencies (e.g. 157 or 325 MHz), 
the LOS mostly sees the inner ring and that gives rise to the MP emission.  The LOS is sufficiently 
further than the outer ring of emission at these frequencies and almost no inter pulse emission 
is observed.  As the frequency increases (e.g. 610 or 1060 MHz), one starts seeing emission coming 
from the second outer ring as well. As a consequence, the IP emission becomes dominant with 
increasing frequency.

For PSR B0826$-$34 we observe frequent nulling and changes of drift rates which are 
simultaneous for both the inner and outer rings.  Esamdin et al. (2005) applied the method 
of phase tracking to the single pulse data and detected in total 13 drift bands in the 
pulse window. Phase locked relation between the inner and outer rings is evident from 
their results (Fig. 6 of Esamdin et al. (2005)). In addition, we observe significant 
correlation between total energy of the main pulse and the inter pulse.  With increasing 
pulse lag this correlation follows similar trend as the auto correlation of total energy 
of the main pulse or the inter pulse. This indicate that the conditions of the 
magnetosphere are similar for inner and outer rings of emission 
(Bhattacharyya et al. 2008b).

\begin{figure}
\plottwo{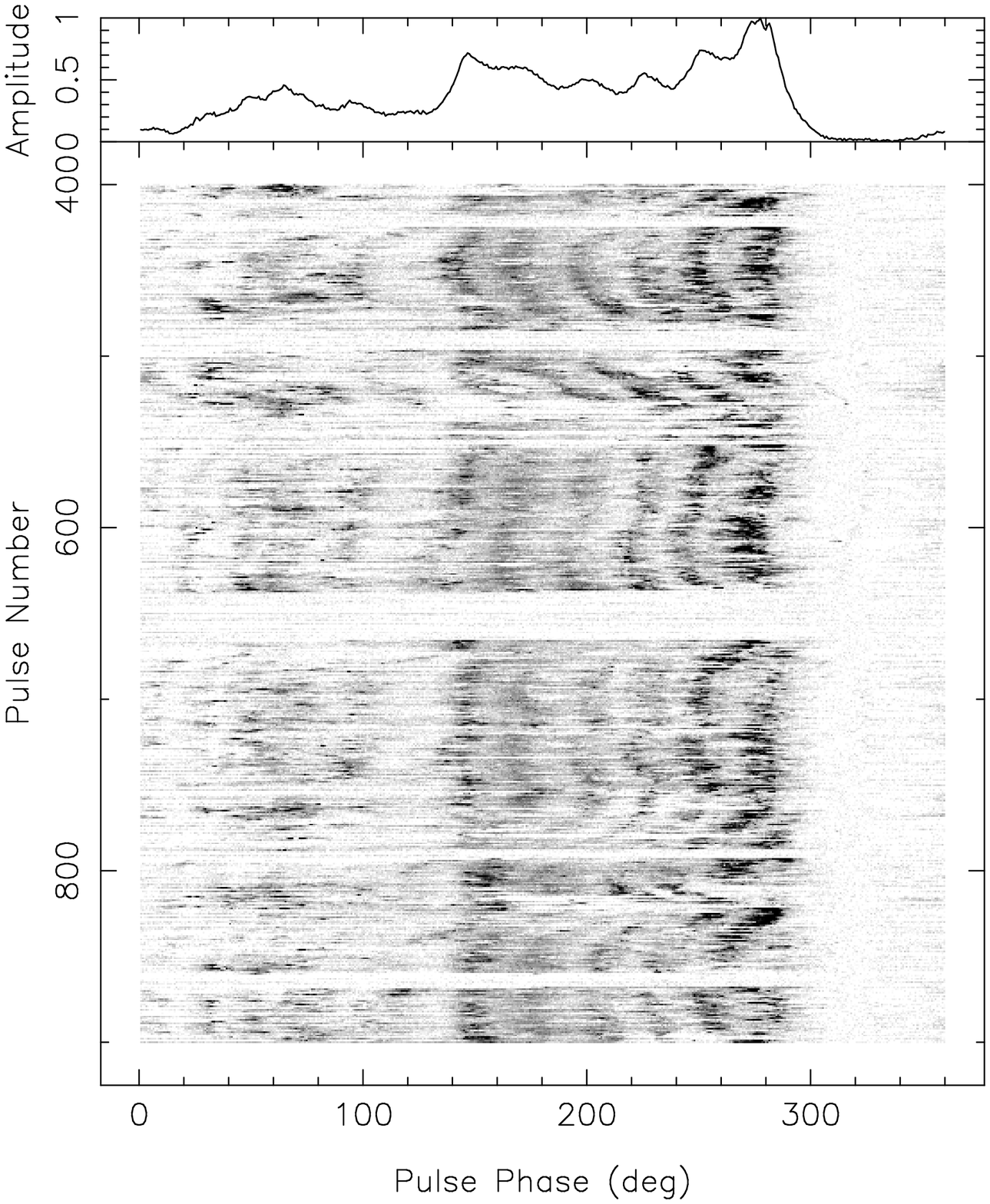}{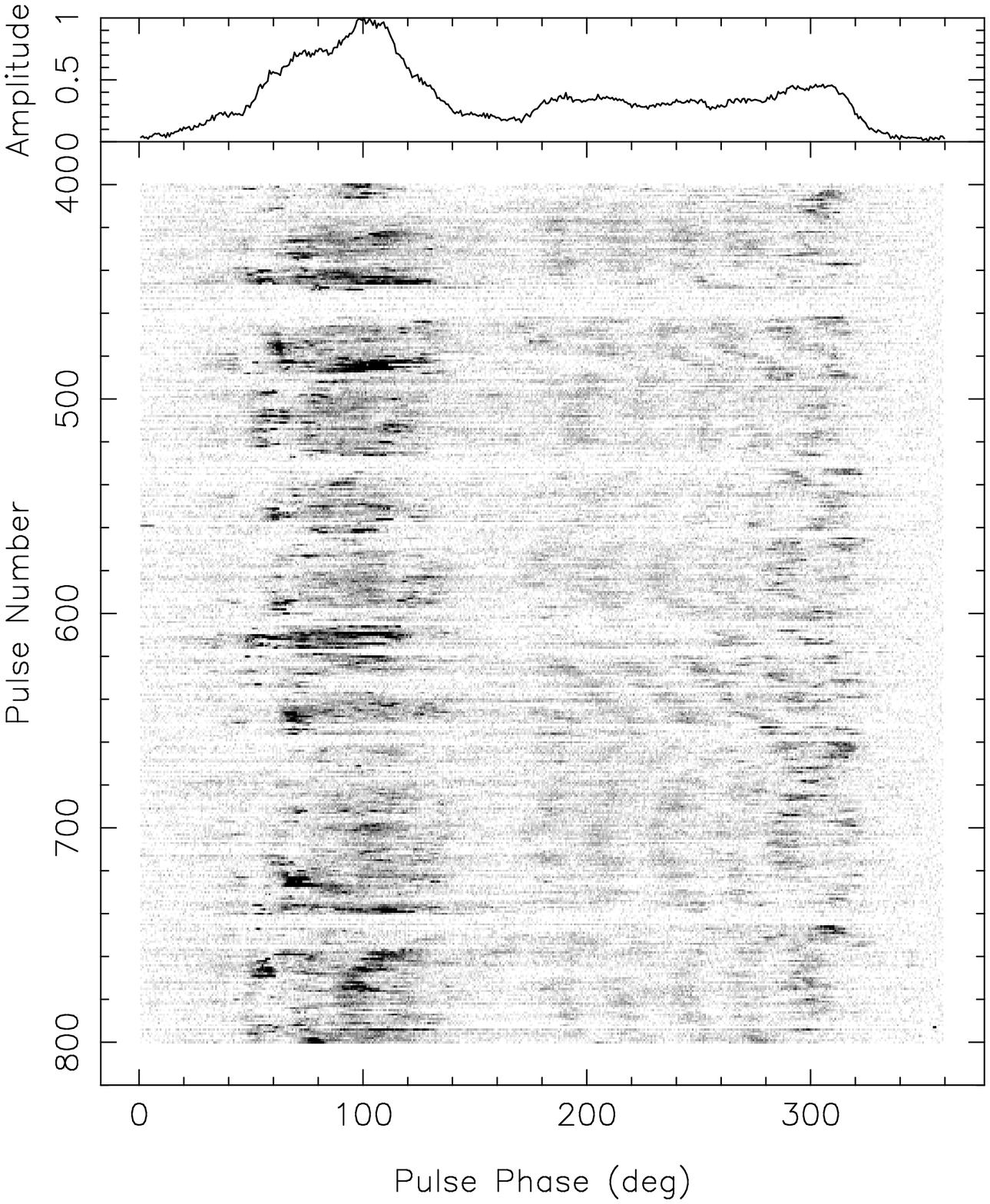}
\caption{{\itshape Left (Fig 2a) : Gray scale plot of the single pulse data from single 
frequency observations of PSR B0826$-$34 at 610 Hz.}
{\itshape Right (Fig 2b) : Same as Fig 2a, but at 1060 Hz.}}
\end{figure}

\section{Conclusions}
We have investigated multiple drift regions in the pulsars B0818$-$41 and B0826$-$34 and
found that the emission from inner and outer rings are locked in phase.  Such a phase 
locked relation between the inner and outer rings could well be a common property 
for all wide profile pulsars, and maybe for all pulsars in general.  The phase locked 
relation implies that inner and outer rings drift with the same angular rate, the
emissions from the two rings are not independent, and conditions responsible for 
drifting are similar in both rings.  This puts constrains on the theoretical models, 
favoring a pan magnetospheric radiation mechanism.\\

Hence our main conclusions are:\\

$\bullet$ Wide profile pulsars provide extra insights into pulsar emission processes;\\

$\bullet$ PSR B0818-41 and PSR B0826$-$34 are wide profile drifting pulsars with 
simultaneous multiple drift regions;\\

$\bullet$ Aliasing can be resolved using the phase relationship between leading and 
trailing outer regions; for PSR B0818$-$41 it indicates P4 ~ 10s, making it the fastest 
known carousel;\\ 

$\bullet$ Phase locked drift between inner and outer rings implies a common electrodynamic 
control over the entire polar cap.\\

%\acknowledgements %%% Text of acknowledgements runs on after this command.

%%% THE BIBLIOGRAPHY
%%%
%%% CONSULT SECTION 3 OF "INSTRUCTIONS FOR AUTHORS" FOR HOW TO USE NATBIB.
%%% AUTHORS ARE ENCOURAGED TO USE EITHER THE "THEBIBLIOGRAPY" ENVIRONMENT
%%% BY UNCOMMENTING (DELETING THE "%" SYMBOL) THE COMMANDS BELOW, OR BY
%%% USING THE BIBTEX ENVIRONMENT. TO FIND OUT WHICH IS APPLICABLE TO YOUR
%%% CONTRIBUTION, CONSULT THE VOLUME EDITORS FOR YOUR PROCEEDINGS.
%%%


\begin{thebibliography}{}
\bibitem[Bhattacharyya et al. (2007)]{Bhattacharyya_etal} Bhattacharyya, B., Gupta, Y., Gil, J., 
Sendyk, M., 2007, {\em MNRAS}, {\bf 377L}, 10B.
\bibitem[Bhattacharyya, Gupta \&  Gil (2008a)]{Bhattacharyya_etal_08} Bhattacharyya, B., Gupta, Y., 
Gil, J., 2008a, {\em MNRAS}, {\bf 383}, 1538B.
\bibitem[Bhattacharyya, Gupta \&  Gil (2008b)]{Bhattacharyya_etal_08b} Bhattacharyya, B., Gupta, Y., 
Gil, J., 2008b, {(\it submitted to MNRAS)} 
\bibitem[]{} {Deshpande}, A.~A., {Rankin}, J.~M., 1999, {\em APJ\/}, {\bf 524}, 1008.
\bibitem[]{} {Gupta}, Y., {Gil}, J., {Kijak}, J., {Sendyk}, M., 2004, {\em A\&A\/},  {\bf 426}, 229.
\bibitem[]{} Qiao, G.~J., Manchester, R.~N., Lyne, A.~G., Gould, D.~M., 1995, {\em   MNRAS\/}, {\bf 274}, 572.
\bibitem[]{} Durdin, J.~M., Large, M.~I., Little, A.~G., Manchester, R.~N., Lyne, A.~G., Taylor, J.~H., 
1979, {\em MNRAS\/}, {\bf 186}, 39P.
\bibitem[]{} Biggs, J.~D., McCulloch, P.~M., Hamilton, P.~A., Manchester, R.~N. \& Lyne, A.~G., 1985, 
{\em MNRAS\/}, {\bf 215}, 281.
\bibitem[]{} Gupta, Y., Gothoskar, P., Joshi, B.~C., Vivekanand, M., Swain, R., Sirothia, S.,
Bhat, N. D.~R., 2000,  {\bf 202}, 277.
\bibitem[]{} Esamdin, A., Lyne, A. G., Graham-Smith, F., Kramer, M.,  Manchester, R. N., Wu, X., 2005, 
{\em MNRAS\/} {\bf 356}, 59.
\end{thebibliography}
\end{document}